\documentclass[aip,jap,preprint,showpacs,amsmath,amssymb]{revtex4-1}

\usepackage{graphicx}
\usepackage{dcolumn}
\usepackage{bm,epstopdf,hyperref}

\begin{document}

\title{A  self-consistent thermodynamic model of metallic systems.\\
Application for the description of gold}

\author{T. Balcerzak}
\email{t$_$balcerzak@uni.lodz.pl}
\author{K. Sza{\l}owski}
\email{kszalowski@uni.lodz.pl}
\affiliation{%
Department of Solid State Physics, Faculty of Physics and Applied Informatics,\\
University of \L\'{o}d\'{z}, ulica Pomorska 149/153, 90-236 \L\'{o}d\'{z}, Poland
}%

\author{M. Ja\v{s}\v{c}ur}
\affiliation{%
Department of Theoretical Physics and Astrophysics, Faculty of Science,\\
P. J. \v{S}af\'arik University, Park Angelinum 9, 041 54 Ko\v{s}ice, Slovak Republic
}%

\date{\today}

\begin{abstract}
A  self-consistent thermodynamic  model of metallic system is presented. The expression for the Gibbs energy is derived, which incorporates elastic (static) energy, vibrational energy  within the Debye model, and electronic part in Hartee-Fock approximation. The elastic energy is introduced by a volume-dependent anharmonic potential. From the Gibbs energy all thermodynamic quantities, as well as the equation of state, are self-consistently obtained. The model is applied for the description of bulk gold in temperature range  $0 \leq T \lesssim 1300$K and external pressure up to 30 GPa. The calculated thermodynamic properties are illustrated in figures and show satisfactory agreement with experimental data. The advantages and opportunities for further development of the method are discussed.
\end{abstract}

\maketitle

\section{Introduction}

The properties of noble metals and, in particular, of gold have intensively been studied, both experimentally \cite{Neighbours, Daniels, Zimmerman, Isaacs, Martin, Cohen, Skelskey, McLean, White, Lynn, Barin, Touloukian, Heinz, Anderson, Collard, Takemura, Yokoo, Kusaba} and theoretically \cite{Hiki, Gupta, Rosen,Sundqvist, Shim, Saxena, Baria, Baria2, Greeff, Godwal, Moriarty, Boettger, Souvatzis, Boettger2, Sun, Garai, Brian, Matsui, Kunc, Jin, Karbasi, Sokolova}.  The studies of gold are connected with its many technical applications, for instance, as a protective coating material, in electronics, as well as in metrology and medicine. For instance, gold provides an accurate pressure  calibration for experiments conducted at high temperatures and pressures \cite{Heinz, Shim, Takemura, Yokoo, Jin, Sokolova}. In recent years, gold nanoparticles attracted much attention as a class of multifunctional materials for biomedical application \cite{Cobley, Bond, Guo}.\\

As far as the experimental studies of  gold are concerned, they include equation of state as well as the lattice and thermodynamic properties, both in low-temperature regime \cite{Zimmerman, Isaacs, Martin, McLean} and in high temperatures and pressures as well \cite{Neighbours, Daniels, Skelskey, Takemura, Kusaba, Yokoo, Matsui}. At the same time, the thermodynamic and vibrational properties have been studied by the theoretical methods, including specific heat calculations \cite{Gupta, Baria2}, phonon dispersion relation \cite{Brian, Lynn}, interatomic interaction and density of states \cite{Baria}, Debye temperatures \cite{Rosen, Baria2}, Gr\"uneisen parameter \cite{Hiki}, lattice constants \cite{Kunc, Matsui} and equation of state \cite{Anderson, Shim, Saxena, Godwal, Moriarty, Boettger, Souvatzis, Boettger2, Sun}. The theoretical methods include studies of anisotropic-continuum model \cite{Hiki}, classical Mie-Gr\"uneisen approach and Birch-Murnaghan equation of state \cite{Anderson, Yokoo}, 
pseudopotential \cite{Baria} and embedded-atom \cite{Brian} 
models, quasiharmonic \cite{Hiki} and anharmomic \cite{Rosen} approach, as well as first principles \cite{Godwal,Moriarty,Boettger, Souvatzis} and density functional theory \cite{Kunc} calculations.\\

Most of the methods mentioned above are very specialized and aimed at characterizing only some of thermodynamic properties, for instance, equation of state, but not at obtaining a full and self-consistent thermodynamic description. Only in few papers an attempt has been made to construct the full thermodynamic description based on the expression for the free energy.  The thermodynamic potential considered there contains the elastic (static) energy, vibrational energy, as well as the electronic contribution \cite{Greeff, Souvatzis}. In this context it is worth mentioning that recently the computational method for the Gibbs energy has been presented, which is suitable for solids under the quasi-harmonic approximation \cite{Otero-de-la-Roza}. In all these papers the electronic energy is taken in the simplest (Sommerfeld) approximation, and the vibrational energy is calculated from the Debye model in quasi-harmonic approximation. On the other hand, the elastic (static) energy calculations involve more elaborate methods, based on 
the density functional theory (DFT).\\

We are aware that DFT methods are not always accessible (and convenient) since they require specialized software and sufficient computational resources. Therefore, apart from referring to such methods it is useful to have also the analytical description of the  system in question. Such description should be based on the Gibbs energy, so giving correct thermodynamic relationships, and being uncomplicated enough for practical use.  In particular, the analytical form of equation of state (EOS) is very useful for understanding mutual relations between physical parameters, as well as contributions to the total pressure from various energy components.  Such EOS should be obtained by demanding minimization of the Gibbs energy functional whereas the system is in equilibrium.\\

Taking the above needs into account, in the present paper we developed a simple approach, based on the derivation of the Gibbs energy, which yields self-consistent thermodynamic description  of  metallic system using analytical solutions. The Gibbs energy consists of Helmholtz free-energy and grand potential ($pV$) term. In turn, the Helmholtz free-energy is constructed from the elastic (static) energy, vibrational energy, and electronic part.
The elastic energy is described by the method presented in Ref.~\cite{Balcerzak}, which is based on the expansion of anharmonic potential in the power series with respect to the volume deformation. The vibrational energy is taken within the Debye model in the analytical approximation, which has been extended for high temperatures. In turn, the electronic energy, apart from the kinetic (Sommerfeld) term, has been completed by the ground state energy, as well as by the exchange energy in Hartree-Fock approximation. From the Gibbs energy the full thermodynamic description is obtained and presented in a form of  analytical expressions. In particular, new EOS has been found for arbitrary temperature $T$ and external pressure $p$. However, the price for the simplicity of the method is that some of initial parameters, defined for $p=0$ and $T=0$, should be taken from experimental data.\\

The formalism has been presented in the Theoretical model Section in detail and then applied for bulk gold in the Numerical results and discussion Section. The numerical calculations have been performed in the range of temperatures from 0 K up to the melting point, and external pressure up to 30 GPa. The calculated thermodynamic properties, for instance,  specific heat, compressibility and thermal expansion, have been presented in figures and compared with experimental data. Some advantages and weak points of the method have been discussed.\\

\section{Theoretical model}
\subsection{General formulation}

The Gibbs free energy of a metallic system is assumed in the form of:\\
\begin{equation}
\label{eq:eq1}
G=F_{\varepsilon}+F_{\rm D}+F_{el}+pV,
\end{equation}
where $F_{\varepsilon}$ is the elastic energy without lattice vibrations, $F_{\rm D}$ is the vibrational energy in Debye approximation and $F_{el}$ is the energy of electronic subsystem.
The elastic energy can be presented in a form of a power series:
\begin{equation}
\label{eq:eq2}
F_{\varepsilon}= N\left( A\varepsilon+\frac{1}{2}B\varepsilon^{2}+\frac{1}{3!}C\varepsilon^{3}+\frac{1}{4!}D\varepsilon^{4}+\frac{1}{5!}E\varepsilon^{5}+\ldots \right)
\end{equation}
where $\varepsilon$ is a parameter characterizing volume deformation, $A,B,C, ...$ are the coefficients and $N$ is the number of atoms in the lattice. The volume deformation is defined by $V=V_0\left(1+\varepsilon \right)$, where $V_0$ is the volume at absolute zero temperature ($T=0$) and in vacuum conditions ($p=0$). This energy is a source of static pressure:
\begin{equation}
\label{eq:eq3}
p_{\varepsilon}=-\left(\frac{\partial F_{\varepsilon}}{\partial V}\right)_T=-\frac{N}{V_{0}}\left(A+B\varepsilon+\frac{1}{2}C\varepsilon^{2}+\frac{1}{3!}D\varepsilon^{3}+\frac{1}{4!}E\varepsilon^{4}+\ldots\right).
\end{equation}
The vibrational energy is taken in the Debye approximation \cite{Wallace}:
\begin{equation}
\label{eq:eq4}
F_{\rm D}=N\left[\frac{9}{8}k_{\rm B}T_{\rm D}+9k_{\rm B}T\left(\frac{T}{T_{\rm D}}\right)^3\,\int_{0}^{y_{\rm D}}y^2 \ln \left(1-e^{-y}\right) dy\right]
\end{equation}
where 
\begin{equation}
\label{eq:eq5}
y_{\rm D}=T_{\rm D}/T
\end{equation}
and $T_{\rm D}$ is the Debye temperature.
Eq.~(\ref{eq:eq4}) can be transformed into a more convenient form:
\begin{equation}
\label{eq:eq6}
F_{\rm D}=N\left[\frac{9}{8}k_{\rm B}T_{\rm D}+3k_{\rm B}T\ln \left(1-e^{-y_{\rm D}}\right)-3k_{\rm B}T\frac{1}{y_{\rm D}^3}\,\int_{0}^{y_{\rm D}}\frac{y^3}{e^{y}-1}dy\right].
\end{equation}
In further considerations we introduce the lattice Gr\"uneisen parameter $\gamma_{\rm D}$ in the following way 
\cite{Dorogokupets1,Dorogokupets2}:
\begin{equation}
\label{eq:eq7}
\gamma_{\rm D}=\gamma_{\rm D}^0\left(V/V_0\right)^q=\gamma_{\rm D}^0\left(1+\varepsilon \right)^q,
\end{equation}
where $q$ is a constant parameter. The Debye temperature $T_{\rm D}$
is connected with the Gr\"uneisen parameter by the relationship \cite{Matsui}:
\begin{equation}
\label{eq:eq8}
T_{\rm D}=T_{\rm D}^0 e^{\left(\gamma_{\rm D}^0-\gamma_{\rm D}\right)/q}\,g(T)=T_{\rm D}^0\, f(\varepsilon )\,g(T)
\end{equation}
where:
\begin{equation}
\label{eq:eq9}
f(\varepsilon )=e^{\left(\gamma_{\rm D}^0-\gamma_{\rm D}\right)/q}= e^{\gamma_{\rm D}^0\left[1-\left(1+\varepsilon \right)^q\right]/q}.
\end{equation}
$T_{\rm D}^0$ and $\gamma_{\rm D}^0$ are the Debye temperature and Gr\"uneisen parameter, respectively, which are taken at $T=0$ and $p=0$. In Eq.~(\ref{eq:eq8}) we introduced $g(T)$ function which weakly depends on temperature and will be specified latter. This function reflects the fact that the Debye temperature depends not only on the volume, but also on temperature itself. It takes into account the so-called "intrinsic anharmonicity" which leads to higher order terms in thermodynamic functions and has also been discussed in several papers\cite{Oganov2004,Dorogokupets2004,Holzapfel2005,Jacobs2005,Zhang2012}\\
It can be easily checked that the above relationship for $T_{\rm D}$ satisfies the classical Gr\"uneisen assumption \cite{Gruneisen1,Gruneisen2}:
\begin{equation}
\label{eq:eq10}
\gamma_{\rm D}=-\frac{V}{\omega_{\rm D}}\left(\frac{\partial \omega_{\rm D}}{\partial V}\right)_T=-\frac{V}{T_{\rm D}}\left(\frac{\partial T_{\rm D}}{\partial V}\right)_T.
\end{equation}
In the low temperature approximation $y_{\rm D} \to \infty$ and we can make use of the integral: $\int_{0}^{\infty}y^3/(e^{y}-1)dy=\pi^4/15$.
Thus, Eq.~(\ref{eq:eq6}) can be transformed to the form:
\begin{equation}
\label{eq:eq11}
F_{\rm D}=N\left[\frac{9}{8}k_{\rm B}T_{\rm D}-\frac{1}{5}\pi^4 k_{\rm B}T\left(\frac{T}{T_{\rm D}}\right)^3\right].
\end{equation}
The above energy gives the vibrational pressure for low temperatures:
\begin{equation}
\label{eq:eq12}
p_{\rm D}=-\left(\frac{\partial F_{\rm D}}{\partial V}\right)_T=3\frac{N}{V_0}k_{\rm B}T_{\rm D}\gamma_{\rm D}\left[\frac{3}{8}+\frac{1}{5}\pi^4 \left(\frac{T}{T_{\rm D}}\right)^4 \right]\frac{1}{1+\varepsilon}
\end{equation}
where $T_{\rm D}$ is given by Eq.~(\ref{eq:eq8}), and $\left(\partial T_{\rm D}/\partial V\right)_T$ is expressed on the basis of Eq.~(\ref{eq:eq10}).\\
On the other hand, for high temperatures, in Eq.~(\ref{eq:eq6}) we can make use of the following series expansion:
\begin{equation}
\label{eq:eq13}
\frac{y}{e^y-1}=\sum_{k=0}^{\infty}B_k\frac{y^k}{k!}
\end{equation}
where $B_k$ are the Bernoulli numbers:
$B_0=1$, $B_1=-\frac{1}{2}$,  $B_2=\frac{1}{6}$,  $B_3=0$,   $B_4=-\frac{1}{30}$,  $B_5=0$,   $B_6=\frac{1}{42}$,  $B_7=0$,   $B_8=-\frac{1}{30}$,  $B_9=0$,   $B_{10}=\frac{5}{66}$,  $B_{11}=0$,   $B_{12}=-\frac{691}{2730}$, etc.\\
This expansion enables us to calculate the integral in the form of a series:
\begin{equation}
\label{eq:eq14}
\int_{0}^{y_{\rm D}}\frac{y^3}{e^{y}-1}dy=\sum_{k=0}^{\infty}\frac{B_k}{k!}\frac{y_{\rm D}^{k+3}}{k+3}.
\end{equation}
For sufficiently high temperatures, when $y_{\rm D}$ is small and satisfies the condition: $T>\frac{1}{2\pi}T_{\rm D}\, \approx 0.16 \,T_{\rm D}$, the series is convergent \cite{Dubinov}. We found that for practical applications, it is sufficient to include in the series only the terms up to fourth order. 
Thus, using the high temperature expansion, the vibrational free energy is given by:
\begin{eqnarray}
\label{eq:eq15}
F_{\rm D}=N\left\{\frac{9}{8}k_{\rm B}T_{\rm D}+3k_{\rm B}T\ln \left(1-e^{-y_{\rm D}}\right)\nonumber\right.\\ \left.-3k_{\rm B}T\left[\frac{1}{3}B_0+\frac{1}{4}B_1y_{\rm D}+\frac{1}{10}B_2y_{\rm D}^2+\frac{1}{168}B_4y_{\rm D}^4 \right]\right\}.
\end{eqnarray}
Such energy gives the following vibrational pressure for high temperatures:
\begin{equation}
\label{eq:eq16}
p_{\rm D}=\frac{N}{V_0}k_{\rm B}T_{\rm D}\gamma_{\rm D}\left[\frac{3}{2}\tanh^{-1}\left(y_{\rm D}/2\right)-\frac{1}{10}\left(1-\frac{1}{42}y_{\rm D}^2\right)y_{\rm D}\right]\frac{1}{1+\varepsilon}.
\end{equation}
It is worth mentioning that without the second term in the square bracket of Eq.~(\ref{eq:eq16}), the remaining $\tanh^{-1}$ function reproduces the pressure of oscillators in the Einstein model \cite{Balcerzak} with rescaled temperature ($T_{\rm D}\to \Theta$, with $\Theta$ being the Einstein characteristic temperature).\\
The electronic free energy $F_{el}$ in Eq.~(\ref{eq:eq1}) can be presented in approximate form as:
\begin{equation}
\label{eq:eq17}
F_{el}=N_{e}\left[-C_{ex}\sqrt{E_{\rm F}}+\frac{3}{5}E_{\rm F}-\frac{\pi^2}{4}\frac{1}{E_{\rm F}}\left(k_{\rm B}T\right)^2\right],
\end{equation}
where $N_e$ is the number of electrons and $E_{\rm F}$ is the Fermi energy, while the exchange constant $C_{ex}$ is equal to:
\begin{equation}
\label{eq:eq24}
C_{ex}=\frac{3}{2\pi}\frac{e^2}{4\pi \epsilon_0}\frac{\sqrt{2m}}{\hbar}.
\end{equation}
The first term in Eq.~(\ref{eq:eq17}) corresponds to the exchange energy in the Hartree-Fock approximation \cite{Suffczynski}. The second term is the kinetic energy for $T=0$, and the last term describes kinetic energy in the low-temperature region, where $T<<T_{\rm F}=E_{\rm F}/k_{\rm B}$, $T_{\rm F}$ denoting the Fermi temperature. The Fermi energy can be presented as a function of the volume:
\begin{equation}
\label{eq:eq18}
E_{\rm F}=E_{\rm F}^0\frac{1}{\left(1+\varepsilon \right)^{\gamma_{\rm F}}}
\end{equation}
where
\begin{equation}
\label{eq:eq19}
E_{\rm F}^0=\frac{\hbar^2}{2m}\left[3\pi^2\frac{N_e}{V_0} \right]^{2/3}
\end{equation}
is the Fermi energy at $T=0$ and $p=0$. In analogy to the  Gr\"uneisen assumption for the lattice parameter (see Eq.~(\ref{eq:eq10})), $\gamma_{\rm F}$-exponent satisfies the equation:
\begin{equation}
\label{eq:eq20}
\gamma_{\rm F}=-\frac{V}{E_{\rm F}}\left(\frac{\partial E_{\rm F}}{\partial V}\right)_T.
\end{equation}
In this paper we assume that $\gamma_{\rm F}$ is a constant parameter (contrary to $\gamma_{\rm D}$, which is volume dependent). 
This parameter can be related to the so-called electronic Gr\"uneisen parameter $\gamma_e$ defined by the formula \cite{White}:
\begin{equation}
\label{eq:eq21}
\gamma_e=-\left(\frac{\partial \ln \textrm{DOS}(E_{\rm F})}{\partial \ln V}\right)_T
\end{equation}
where $ \textrm{DOS}(E_{\rm F})$ is the density of states at Fermi surface. From Eqs.~(\ref{eq:eq20}) and (\ref{eq:eq21}) one can obtain the relationship:
\begin{equation}
\label{eq:eq22}
\gamma_e=1-\frac{\gamma_{\rm F}}{2}.
\end{equation}
For $\gamma_{\rm F}=2/3$, which is the free-electron case, we obtain $\gamma_e=\gamma_{\rm F}$.\\
From the expression (\ref{eq:eq17}) the electronic part of the pressure can be found as:
\begin{equation}
\label{eq:eq23}
p_{el}=-\left(\frac{\partial F_{el}}{\partial V}\right)_T=\frac{N_e}{V_0}\gamma_{\rm F}\left[-\frac{1}{2}C_{ex}\sqrt{E_{\rm F}}+\frac{3}{5}E_{\rm F}+\frac{\pi^2}{4}\frac{1}{E_{\rm F}}\left(k_{\rm B}T\right)^2 \right]\frac{1}{1+\varepsilon},
\end{equation}
which is valid for $T\ll T_{\rm F}$. 
Under the external pressure $p$, the equation of state follows from the equilibrium condition:
\begin{equation}
\label{eq:eq25}
p=p_{\varepsilon}+p_{\rm D}+p_{el},
\end{equation}
where $p_{\varepsilon}$ is given by Eq.~(\ref{eq:eq3}), $p_{el}$ - by Eq.~(\ref{eq:eq23}), and $p_{\rm D}$ is given for the low or high temperature regions by Eqs.~(\ref{eq:eq12}) or (\ref{eq:eq16}), respectively.\\

\subsection{The Gr\"uneisen relationship}

For the systems described by $(p, V, T)$-variables we can make use of the exact thermodynamic relationship:
\begin{equation}
\label{eq:eq26}
\frac{\alpha_p}{\kappa_T} = \left(\frac{\partial p}{\partial T}\right)_V,
\end{equation}
where $\alpha_p$ is the thermal volume expansion coefficient:
\begin{equation}
\label{eq:eq27}
\alpha _p = \frac{1}{V}\left( \frac{\partial V}{\partial T}\right)_{p}
\end{equation}
and $\kappa_T$ is the isothermal compressibility:
\begin{equation}
\label{eq:eq28}
\kappa _T = -\frac{1}{V}\left( \frac{\partial V}{\partial p}\right)_{T}.
\end{equation}
For pressure given in the form of Eq.~(\ref{eq:eq25}) the temperature partial derivatives can be calculated as follows:
\begin{equation}
\label{eq:eq29}
\left( \frac{\partial p_{\varepsilon}}{\partial T}\right)_V=0,
\end{equation}
\begin{equation}
\label{eq:eq30}
\left( \frac{\partial p_{\rm D}}{\partial T}\right)_V=\gamma_{\rm D}\frac{C_V^{\rm D}}{V},
\end{equation}
where $C_V^{\rm D}$ is the phononic heat capacity, and
\begin{equation}
\label{eq:eq31}
\left( \frac {\partial p_{el}}{\partial T}\right)_V=\gamma_{\rm F}\frac{C_V^{el}}{V},
\end{equation}
where $C_V^{el}$ is the electronic heat capacity.
The phononic heat capacity at constant volume for low temperatures $(T\ll T_{\rm D})$ can be found from the relationship:
\begin{eqnarray}
\label{eq:eq32}
C_V^{\rm D}&=&-T\left( \frac{\partial^2 F_{\rm D}}{\partial T^2}\right)_V\\
&=&\frac{12}{5}\pi^4 Nk_{\rm B}\left( \frac {T}{T_{\rm D}}\right)^3w(T)
-9Nk_{\rm B}T_{\rm D}\left[ \frac{1}{8}+\frac{\pi^4}{15}\left(\frac{T}{T_{\rm D}}\right)^4\right]\,\frac{T}{g(T)}\,\frac{\partial^2 g(T)}{\partial T^2}\nonumber
\end{eqnarray}
where $F_{\rm D}$ is taken from Eq.~(\ref{eq:eq11}). The new function $w(T)$ is defined by:
\begin{equation}
\label{eq:eq33}
w(T)=\left[1-\frac{T}{g(T)}\,\frac{\partial g(T)}{\partial T} \right]^2.
\end{equation}
On the other hand, for the high temperature region we have on the basis of Eq.~(\ref{eq:eq15}):
\begin{eqnarray}
\label{eq:eq34}
C_V^{\rm D}=3Nk_{\rm B}\left(\frac{T_{\rm D}}{T}\right)^2\left[ \frac{e^{T_{\rm D}/T}}{\left( e^{T_{\rm D}/T}-1\right)^2} +\frac{1}{30}-\frac{1}{420}\left( \frac{T_{\rm D}}{T}\right)^2\right]w(T)\nonumber\\
-3Nk_{\rm B}T_{\rm D}\left[\frac{1}{e^{T_{\rm D}/T}-1}+\frac{1}{2}-\frac{1}{30}\, \frac{T_{\rm D}}{T}+\frac{1}{1260}\left(\frac{T_{\rm D}}{T}\right)^3\right]
\,\frac{T}{g(T)}\,\frac{\partial^2 g(T)}{\partial T^2}.
\end{eqnarray}
In further calculations we will assume that $g(T)$ function only weakly depends on temperature and has the simple linear form:
\begin{equation}
\label{eq:eq35}
g(T)=1+r\frac{T}{T_{\rm D}^0}
\end{equation}
where $r$ is a constant parameter ($r\ll 1$) over the whole temperature region. Then, in right-hand side of Eqs.~(\ref{eq:eq32}) and (\ref{eq:eq34}) the second terms containing $\partial^2 g(T)/\partial T^2$ vanish, whereas in the first terms $w(T)$ takes the form of:
\begin{equation}
\label{eq:eq36}
w(T)=\left(1-\frac{rT/T_{\rm D}^0}{1+rT/T_{\rm D}^0}\right)^2.
\end{equation}
We see that $w(T)$ function for $r>0$ can enforce some decrease of the specific heat vs. temperature in comparison with the case when $g(T)=const$. We found that such a possibility can be useful in order to reproduce better the experimental data.\\

The electronic heat capacity at constant volume is given by the formula:
\begin{equation}
\label{eq:eq37}
C_V^{el}=-T\left(\frac{\partial^2 F_{el}}{\partial T^2}\right)_V=N_e\frac{ \pi^2}{2}\frac{1}{E_{\rm F}}k_{\rm B}^2T
\end{equation}
which is valid for low temperatures $(T\ll T_{\rm F})$ in the electronic scale.\\
Substituting the pressure derivatives (Eqs.~(\ref{eq:eq29}-\ref{eq:eq31})) into Eq.~(\ref{eq:eq26}) we obtain the Gr\"uneisen relationship for complex (phononic and electronic) system which can be presented as:
\begin{equation}
\label{eq:eq38}
\frac{\alpha_p}{\kappa_T} = \frac{\gamma_{\rm D}C_V^{\rm D}+\gamma_{\rm F}C_V^{el}}{V}=\frac{\gamma^{\rm eff}}{V}C_V
\end{equation}\\
where
\begin {equation}
\label{eq:eq39}
C_V=C_V^{\rm D}+C_V^{el},
\end{equation}
and
\begin {equation}
\label{eq:eq40}
\gamma^{\rm eff}=\frac{\gamma_{\rm D}C_V^{\rm D}+\gamma_{\rm F}C_V^{el}}{C_V^{\rm D}+C_V^{el}}.
\end{equation}
Therefore, $\gamma^{\rm eff}$ is the effective Gr\"uneisen parameter for the electronic and phononic complex system.\\

\subsection{The dimensionless equation of state (EOS)}

Equation of state (\ref{eq:eq25}) can be presented in a dimensionless form which is convenient for numerical calculations. First, we introduce the reference energy $A_{\rm D}^0$ for normalization of various energy coefficients. We define:
\begin{equation}
\label{eq:eq41}
A_{\rm D}^0=k_{\rm B}T_{\rm D}^0,
\end{equation}
where $T_{\rm D}^0$ is the Debye temperature at 0K.
Then, we can introduce the dimensionless pressure:
\begin{equation}
\label{eq:eq42}
\tilde p = \frac{V_0}{N}\frac{1}{A_{\rm D}^0}p,
\end{equation}
and dimensionless temperature:
\begin{equation}
\label{eq:eq43}
\tilde T = \frac{T}{T_{\rm D}^0}.
\end{equation}
By the same token, the elastic constants can be normalised as follows:
\begin{equation}
\label{eq:eq44}
\tilde A = \frac{A}{A_{\rm D}^0}, \; \tilde B=\frac{B}{A_{\rm D}^0}, \; \tilde C=\frac{C}{A_{\rm D}^0}, \; \tilde D=\frac{D}{A_{\rm D}^0}, \; \tilde E=\frac{E}{A_{\rm D}^0}.
\end{equation}
Similarly, the dimensionless Fermi and exchange energies can be written as
\begin{equation}
\label{eq:eq45}
\tilde {E_{\rm F}^0} = \frac{E_{\rm F}^0}{A_{\rm D}^0}
\end{equation}
and
\begin{equation}
\label{eq:eq46}
\tilde {E_{ex}^0} = -\frac{C_{ex} \sqrt{E_{\rm F}^0}}{A_{\rm D}^0},
\end{equation}
respectively.\\
With the above notation the equation of state (Eq.~(\ref{eq:eq25})) in the low temperature limit can be presented in the dimensionless form:
\begin{eqnarray}
\label{eq:eq47}
\tilde p +\tilde A +\tilde B\varepsilon+\frac{1}{2}\tilde C\varepsilon^2+\frac{1}{3!}\tilde D\varepsilon^3+\frac{1}{4!}\tilde E\varepsilon^4=\nonumber\\
3\gamma_{\rm D}f(\varepsilon)g(T)\left[\frac{3}{8}+\frac{1}{5}\pi^4 \left(\frac{\tilde T}{f(\varepsilon)g(T)}\right)^4\right]\frac{1}{1+\varepsilon}\nonumber\\
+\frac{N_e}{N}\gamma_{\rm F}\left[\frac{1}{2}\tilde{E_{ex}^0}\frac{1}{\left(1+\varepsilon \right)^{\gamma_{\rm F}/2}}+\frac{3}{5}\tilde {E_{\rm F}^0}\frac{1}{\left(1+\varepsilon \right)^{\gamma_{\rm F}}}+\frac{\pi^2}{4}\frac{\tilde T^2}{\tilde {E_{\rm F}^0}}\left(1+\varepsilon \right)^{\gamma_{\rm F}}\right]\frac{1}{1+\varepsilon}.
\end{eqnarray}
From the equilibrium condition for $\tilde T =0$ and $\tilde p=0$ we must have $\varepsilon =0$. This condition allows to determine the $\tilde A$-coefficient, namely:
\begin{equation}
\label{eq:eq48}
\tilde A=\frac{9}{8}\gamma_{\rm D}^0+\frac{N_e}{N}\gamma_{\rm F}\left[\frac{1}{2}\tilde {E_{ex}^0}+\frac{3}{5}\tilde {E_{\rm F}^0}\right].
\end{equation}
Now, substituting Eq.~(\ref{eq:eq48}) into (\ref{eq:eq47}) we finally obtain the equation of state in the low temperature region as:
\begin{eqnarray}
\label{eq:eq49}
\tilde p +\frac{9}{8}\gamma_{\rm D}^0 +\tilde B\varepsilon+\frac{1}{2}\tilde C\varepsilon^2+\frac{1}{3!}\tilde D\varepsilon^3+\frac{1}{4!}\tilde E\varepsilon^4+\frac{1}{2}\frac{N_e}{N}\gamma_{\rm F}\tilde{E_{ex}^0}\left[1-\frac{1}{\left(1+\varepsilon \right)^{\gamma_{\rm F}/2+1}}\right]=\nonumber\\
3\gamma_{\rm D}\, f(\varepsilon)g(T)\left[\frac{3}{8}+\frac{1}{5}\pi^4 \left(\frac{\tilde T}{f(\varepsilon)g(T)}\right)^4\right]\frac{1}{1+\varepsilon }\nonumber\\
+\frac{3}{5}\frac{N_e}{N}\gamma_{\rm F}\tilde {E_{\rm F}^0}\left[ \frac{1}{\left(1+\varepsilon \right)^{\gamma_{\rm F}+1}} -1\right]+\frac{N_e}{N}\frac{\pi^2}{4}\gamma_{\rm F}\frac{\tilde T^2}{\tilde {E_{\rm F}^0}}\left(1+\varepsilon \right)^{\gamma_{\rm F}-1}.
\end{eqnarray}
When $T\to 0$, this equation can be linearized with respect to $\varepsilon$. Then, by comparison with the approximate formula $\varepsilon \approx -\kappa_0\frac{NA_{\rm D}^0}{V_0}\tilde p\,$, which is valid in the limit $T\to 0$, the $\tilde B$ coefficient can be determined:
\begin{equation}
\label{eq:eq50}
\tilde B=\frac{V_0}{N}\frac{1}{A_{\rm D}^0 \kappa_0}-\frac{9}{8}\gamma_{\rm D}^0\left(\gamma_{\rm D}^0-q+1\right)-\frac{3}{5}\gamma_{\rm F}\left(\gamma_{\rm F}+1\right)\frac{N_e}{N}\tilde {E_{\rm F}^0}-\frac{1}{2}\gamma_{\rm F}\left(\frac{\gamma_{\rm F}}{2}+1\right)\frac{N_e}{N}\tilde {E_{ex}^0},
\end{equation}
where $\kappa_0$ is the isothermal compressibility at $T=0$ and $p=0$. All parameters necessary to calculate $\tilde B$ from Eq.~(\ref{eq:eq50}) can be taken from experimental data.\\

In the high temperature region the vibrational pressure $p_{\rm D}$ should be taken from Eq.~(\ref{eq:eq16}) instead of  Eq.~(\ref{eq:eq12}). This leads to the dimensionless EOS for high temperatures:
\begin{eqnarray}
\label{eq:eq51}
\tilde p +\frac{9}{8}\gamma_{\rm D}^0 +\tilde B\varepsilon+\frac{1}{2}\tilde C\varepsilon^2+\frac{1}{3!}\tilde D\varepsilon^3+\frac{1}{4!}\tilde E\varepsilon^4+\frac{1}{2}\frac{N_e}{N}\gamma_{\rm F}\tilde{E_{ex}^0}\left[1-\frac{1}{\left(1+\varepsilon \right)^{\gamma_{\rm F}/2+1}}\right]=\nonumber\\
\frac{\gamma_{\rm D} f(\varepsilon)g(T)}{1+\varepsilon}\left[\frac{3}{2}\tanh^{-1}\left(\frac{f(\varepsilon )g(T)}{2\tilde T}\right)-\frac{1}{10}\frac{f(\varepsilon )g(T)}{\tilde T}+\frac{1}{420} \left( \frac{f(\varepsilon )g(T)}{\tilde T}\right)^3\right]
\nonumber\\
+\frac{3}{5}\frac{N_e}{N}\gamma_{\rm F}\tilde {E_{\rm F}^0}\left[ \frac{1}{\left(1+\varepsilon \right)^{\gamma_{\rm F}+1}} -1\right]+\frac{N_e}{N}\frac{\pi^2}{4}\gamma_{\rm F}\frac{\tilde T^2}{\tilde {E_{\rm F}^0}}\left(1+\varepsilon \right)^{\gamma_{\rm F}-1}
\end{eqnarray}
By comparison of the Eqs.~(\ref{eq:eq51}) and (\ref{eq:eq49}) one can see that only the phononic part on the r.h.s. of these equation has been modified.\\

\subsection{Calculation of other thermodynamic quantities}

From the equation of state, the isotherms ($\varepsilon(p)$ for $T=const.$) or isobars ($\varepsilon(T)$ for $p=const.$) can be found directly. Other thermodynamic properties result from differentiation of the equation of state. For instance, the thermal volume expansion coefficient $\alpha_p$ is given by Eq.~(\ref{eq:eq27}), whereas the isothermal compressibility $\kappa_T$ is given by Eq.~(\ref{eq:eq28}). \\
The adiabatic compressibility is measured in the experiment. It is given by the definition:
\begin{equation}
\label{eq:eq52}
\kappa _S = -\frac{1}{V}\left( \frac{\partial V}{\partial p}\right)_{S}
\end{equation}
where the partial derivative is taken at constant entropy $S$, where $S=-\left(\partial G/\partial T\right)_p$.\\
In turn, the heat capacity at constant pressure is defined as:
\begin{equation}
\label{eq:eq53}
C_p=-T\left( \frac{\partial^2 G}{\partial T^2}\right)_p.
\end{equation}
Having calculated $\alpha_p$, $\kappa_T$ and $C_V$ on the basis of EOS, the quantities defined above ($\kappa _S$ and $C_p$) can be conveniently obtained from the exact thermodynamic relationships:
\begin{equation}
\label{eq:eq54}
\kappa_S = \kappa_T/\left( 1+ T V \frac{\alpha_p^2}{C_V \kappa_T}\right) 
\end{equation}
and
\begin{equation}
\label{eq:eq55}
C_p = C_V \left(1+ T V \frac{\alpha_p^2}{C_V \kappa_T} \right).
\end{equation}
With the help of the generalized Gr\"uneisen relationship (Eq.~(\ref{eq:eq38})) the above formulas can also be presented in a more elegant form:
\begin{equation}
\label{eq:eq56}
\kappa_S = \kappa_T / \left( 1+ T \alpha_p \gamma^{\rm eff}\right) 
\end{equation}
and
\begin{equation}
\label{eq:eq57}
C_p = C_V \left(1+ T \alpha_p \gamma^{\rm eff} \right)
\end{equation}
where the effective Gr\"uneisen parameter $\gamma^{\rm eff}$ is defined by Eq.~(\ref{eq:eq40}).\\
It is worth mentioning that using the above equations the exact thermodynamic identity
\begin{equation}
\label{eq:eq58}
\frac{C_p}{C_V}=\frac{\kappa_T}{\kappa_S}
\end{equation}
remains fulfilled.\\
In order to illustrate the method, some exemplary numerical calculations based on the proposed formalism will be presented in the next Section.\\

\section{Numerical results and discussion}
 
\begin{figure}
\includegraphics[scale=1.0]{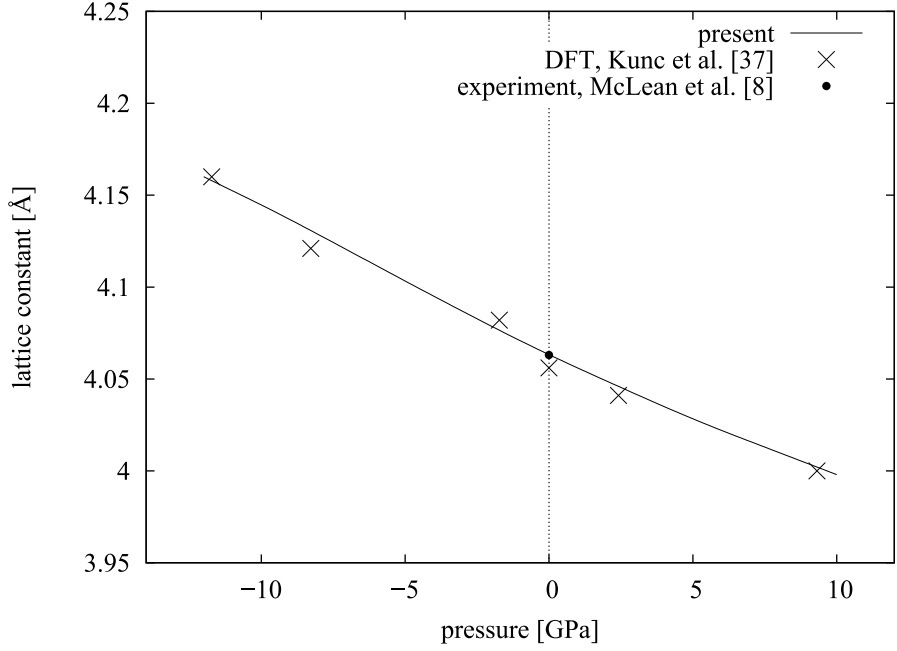}
\caption{Lattice constant ${\it a}$ vs. pressure for $T=0$.}
\label{fig1}
\end{figure} 

As an application of the theory presented in previous Section we shall describe thermodynamic properties of bulk gold. For such metallic system many experimental results are available which can be compared with calculations performed within the present model. Some of the experimental data at 0 K can serve as the input parameters for present formalism. For instance, the Debye temperature at 0 K amounts to $T_{\rm D}^0=164$ K \cite{Zimmerman}. On this basis we can calculate the reference energy $A_{\rm D}^0$ (see Eq.(~\ref{eq:eq41})), namely $A_{\rm D}^0=2.264 \times 10^{-21}$ J. Another parameter is the atomic volume at 0 K, $V_0/N=1.677\times 10^{-29} {\rm m}^3$, which is estimated from Ref.~\cite{McLean}. The isothermal compressibility at 0 K can be calculated as the inverse of the bulk modulus. On the basis of Refs.~\cite{McLean, Neighbours} we obtained the value $\kappa_0=5.546\times 10^{-12} {\rm Pa}^{-1}$.
As far as the Gr\"uneisen coefficient is concerned, we assumed the value of $\gamma_{\rm D}^0=2.95$, which is an average value from the data reported in Refs.~\cite{White} and \cite{McLean}. Then, the parameter $q$ is assumed as $q=0.8$, which is one of the possible values considered in Ref.~\cite{Matsui}. Regarding electronic properties, we assumed that the parameter $\gamma_{\rm F}=2/3= const.$ in the whole temperature region, which corresponds to the free electron model. The electron density per atom is equal to $N_e/N=1$. Taking into account the above input data, we calculated the normalized Fermi energy at 0K as ${\tilde {E_{\rm F}^0}}=394$ and the exchange energy $\tilde{E_{ex}^0}=-105$. Moreover, the $\tilde B$-parameter can be calculated on the basis of Eq.~(\ref{eq:eq50}), and yields the value  $\tilde B=1109.5$.\\

\begin{figure}
\includegraphics[scale=1.0]{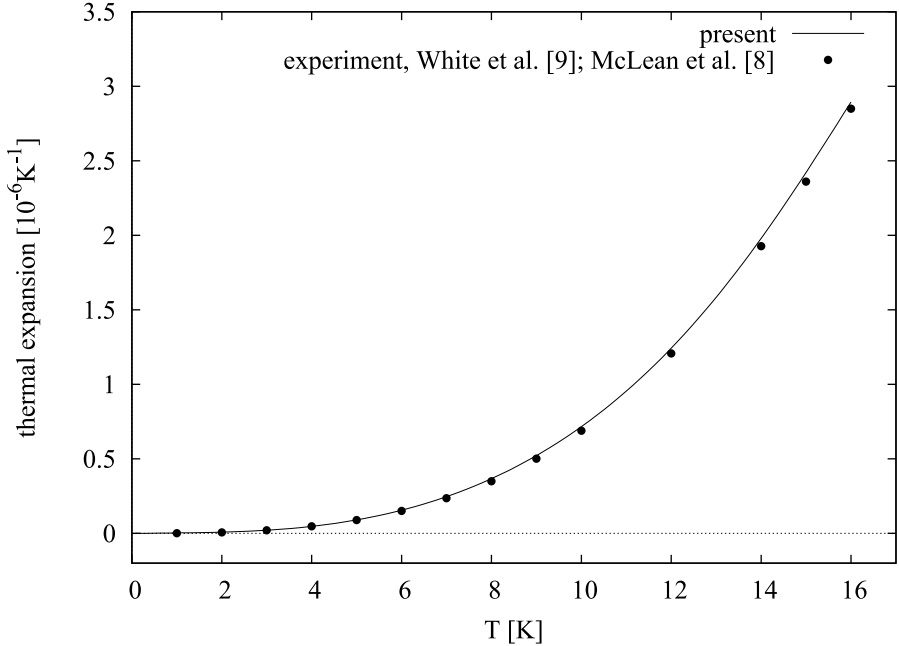}
\caption{Thermal volume expansivity of Au, $\alpha _p$, in low temperature region.}
\label{fig2}
\end{figure}

\begin{figure}
\includegraphics[scale=1.0]{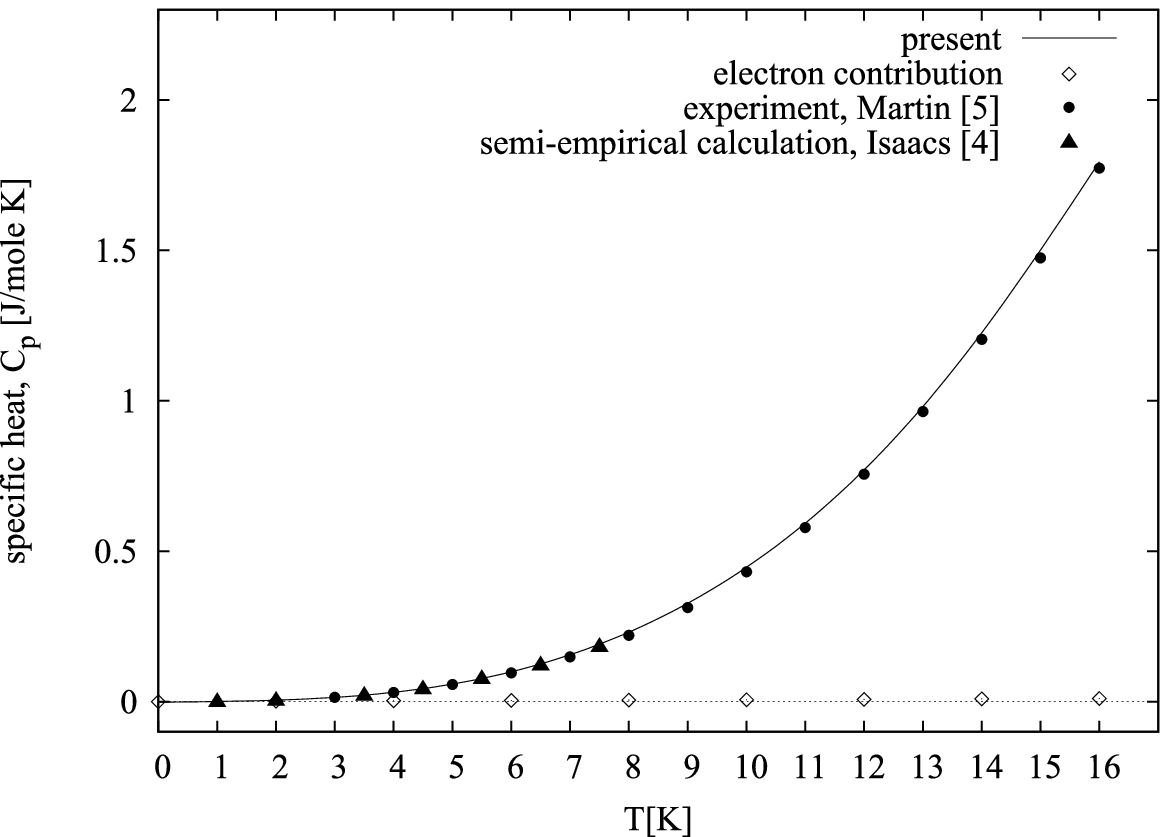}
\caption{Specific heat of Au at constant pressure, $C_p$, in low temperature region. By the diamond symbols the electron contribution to the specific heat is shown.}
\label{fig3}
\end{figure}

Other theoretical parameters, which are necessary for further calculations, are connected with the coefficients in the expression for the elastic energy (Eq.~(\ref{eq:eq2})). They are treated as the fitting parameters in our theory. We found that the best fit is obtained for the following set of coefficients: $\tilde C=-8000$, $\tilde D=200000$, $\tilde E=1100000$ for $\varepsilon <0$, and $\tilde C=-9000$, $\tilde D=5000$, $\tilde E=900000$ for $\varepsilon >0$. These coefficients reflect the asymmetry of the elastic energy with respect to the sign of deformation $\varepsilon$. Finally, the $r$-coefficient in $g(T)$ - function (defined by Eq.~(\ref{eq:eq35})) is assumed as: $r=0.0025$. Having the above set of starting parameters all thermodynamic properties can be calculated for arbitrary $T>0$ and $p>0$. For the gold crystal we explored the temperature range $0 \le T< T_{\rm M}$, where $ T_{\rm M}$ is the melting temperature, and the pressure $p$ was from the range up to several tens of GPa.\\

\begin{figure}
\includegraphics[scale=1.0]{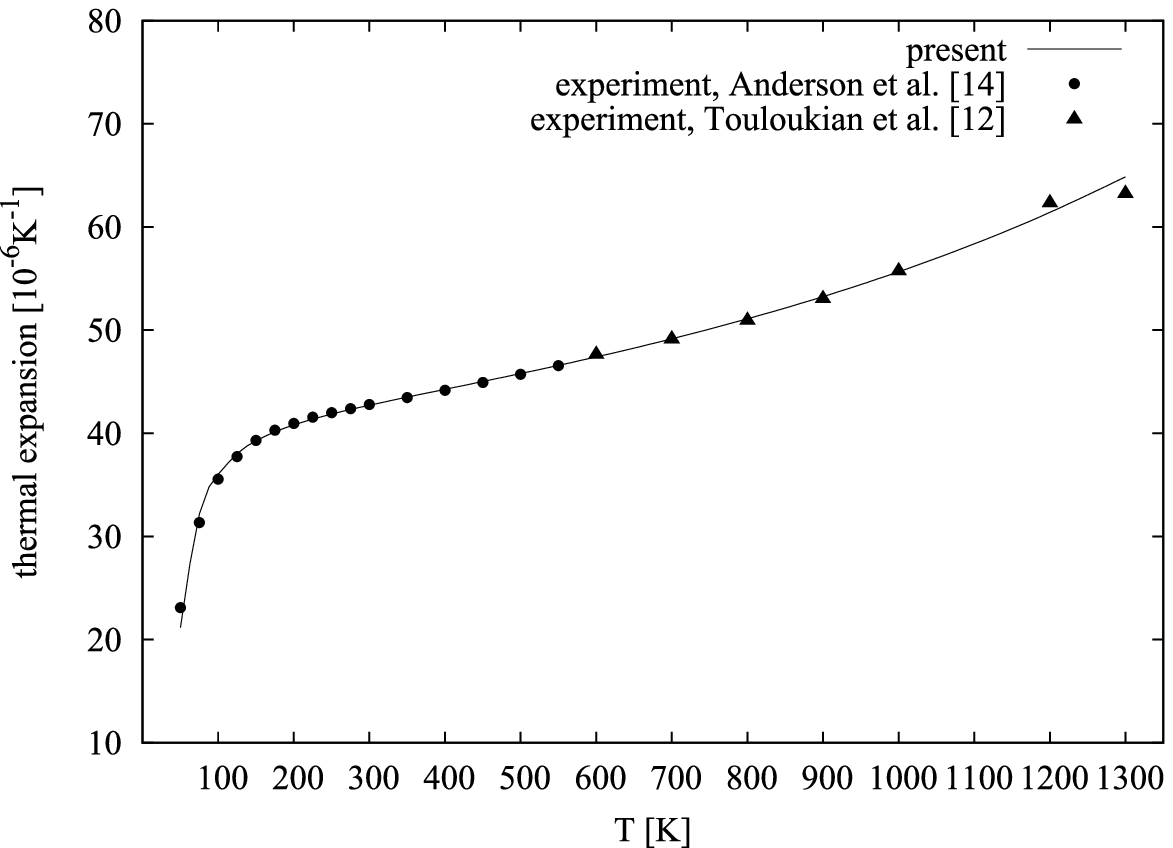}
\caption{Thermal volume expansivity of Au, $\alpha _p$, vs. temperature ($p$=0).}
\label{fig4}
\end{figure}

\begin{figure}
\includegraphics[scale=1.0]{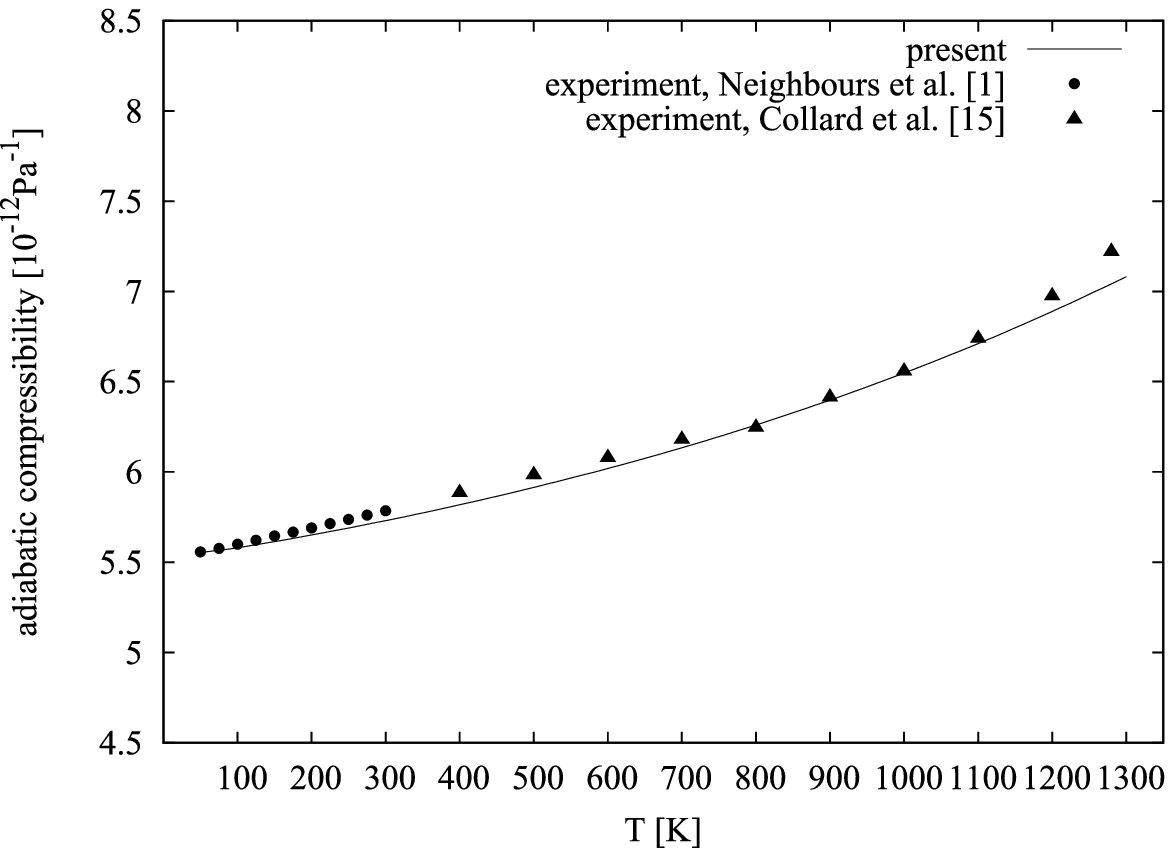}
\caption{Adiabatic compressibility of Au, $\kappa_S$, vs. temperature (ambient pressure).}
\label{fig5}
\end{figure}

We start our calculations with the low-temperature region. In particular, for $T=0$, on the basis of EOS the volume dependence vs. pressure can easily be obtained. Hence, the lattice constant $a$ vs. $p$ can be calculated. The result is presented in Fig.~\ref {fig1}. We see from that figure that the present result fits well DFT (LDA) calculations \cite{Kunc}. It also agrees with the experimental point for $p=0$ and $T=0$, namely $a_0=4.06 {\rm \AA}$ \cite{McLean}. It is worth mentioning that the slope of the curve at ($p=0$, $T=0$) can be related to the isothermal compressibility $\kappa _0$ and well reproduces the experimental value $\kappa_0=5.546\times 10^{-12} {\rm Pa}^{-1}$ \cite{McLean}.\\

In Fig.~\ref{fig2} the thermal volume expansivity of Au, $\alpha_p$ is plotted vs. $T$ in the low-temperature region. The results are compared with the experimental data which have been averaged from two sources: Ref.~\cite{White}  and Ref.~\cite{McLean}. An excellent agreement of numerical results with the experimental points is seen.\\

\begin{figure}
\includegraphics[scale=1.0]{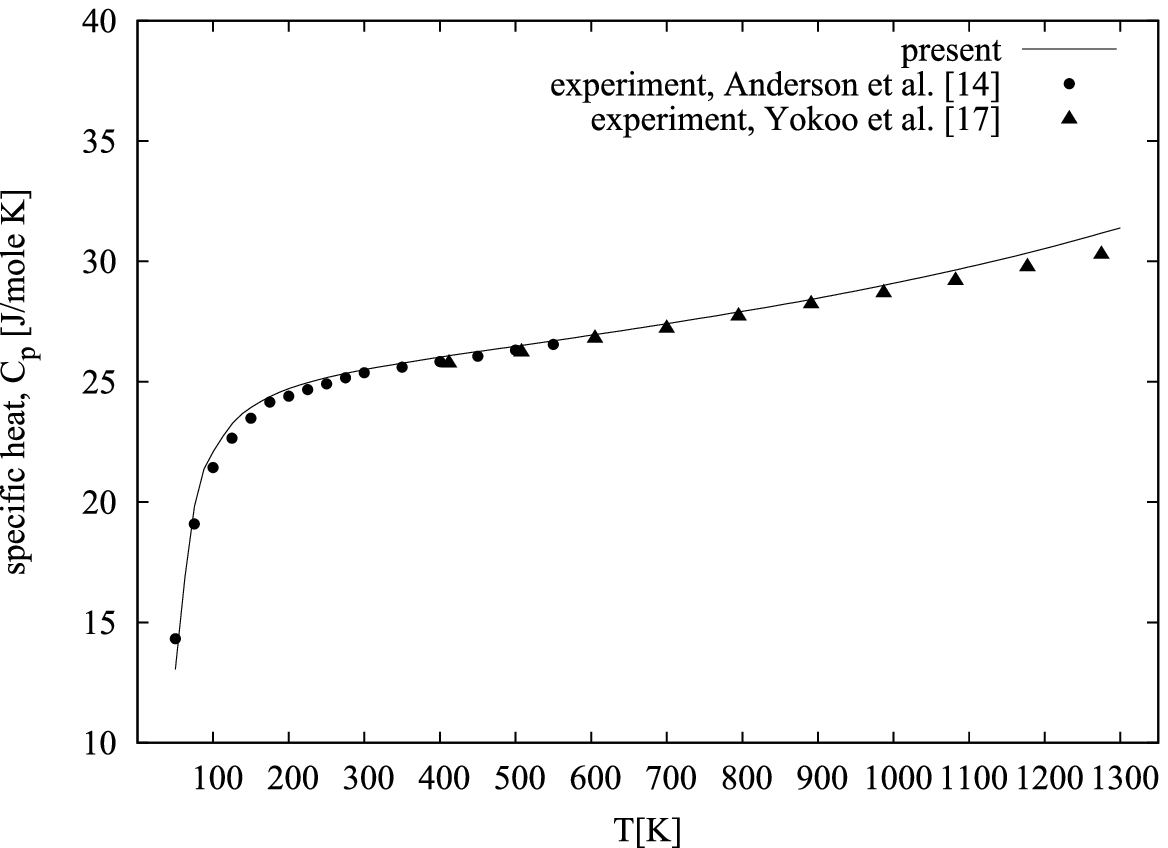}
\caption{Specific heat of Au at constant pressure, $C_p$, vs. temperature.}
\label{fig6}
\end{figure}

\begin{figure}
\includegraphics[scale=1.0]{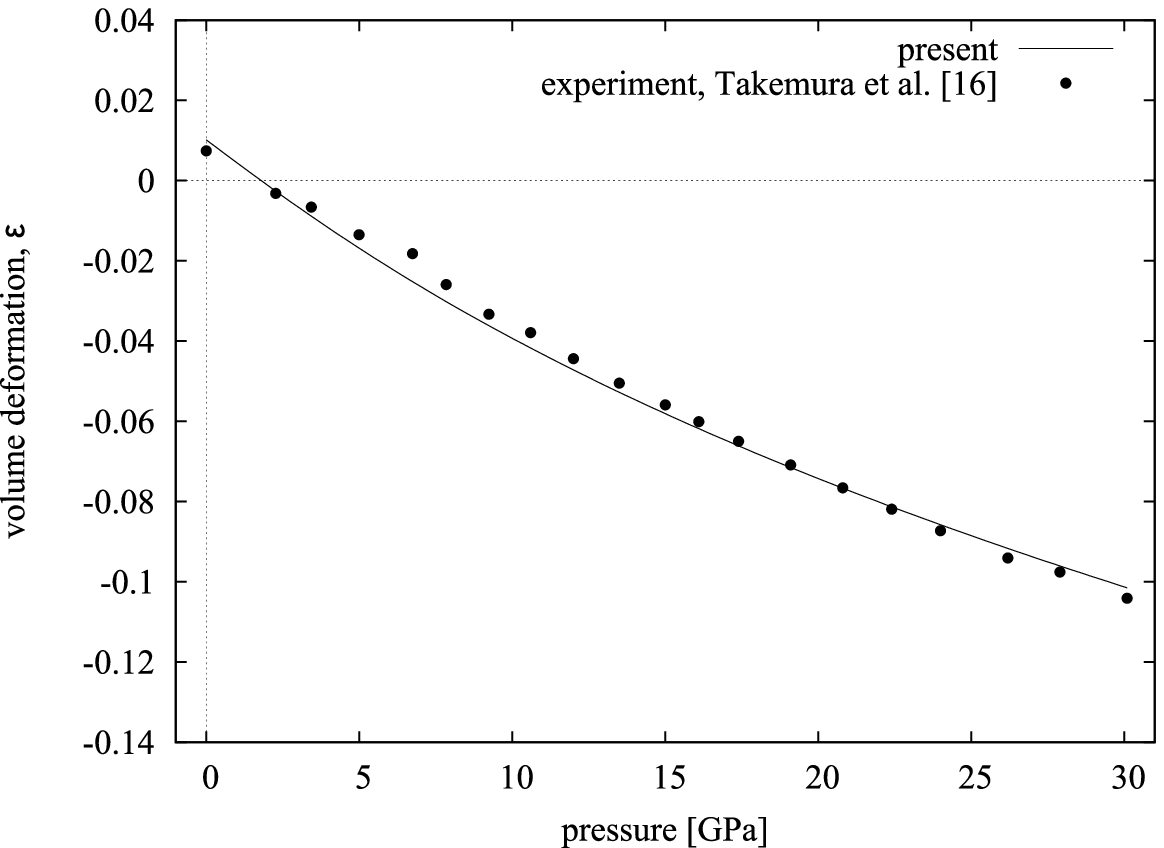}
\caption{Isotherm of Au (volume deformation $\varepsilon$ vs. pressure $p$) for $T$=300K.}
\label{fig7}
\end{figure}

In Fig.~\ref{fig3} the specific heat at constant pressure is presented vs. temperature in the same low-temperature region ($T\lesssim 0.1\, T_{\rm D}^0$). The present result is compared  with the experimental data from Ref.~\cite{Martin}.  We also present there the results of calculations based on the semi-empirical formula:  $C_p(T)=0.729\times T+0.4504\times T^3  - 0.00048\times T^5$. This formula has been  found in Ref.~\cite{Isaacs} and reflects the anomalous behaviour of the specific heat (i.e., the negative coefficient at $\sim T^5$ term) in very low temperature region. From Fig.~\ref{fig3} one can conclude  that the agreement of the theoretical results with the experimental data is very satisfactory. In this figure, by the diamond symbols the electron contribution to the calculated specific heat is also shown. This contribution is very small in the range of intermediate temperatures, however, it exceeds the phononic specific heat in the extremely low temperature region, namely when $T \le 1.2$ K (at $T=1.2$ K, $C_V^{el}=C_V^{\rm D}\approx 7.6 \times 10^{-4}$J/mole K). On the other hand, at $T=16$ K the electron contribution constitutes only 0.56\% of the total specific heat. Due to linear increase vs. temperature the electron fraction of the specific heat will increase again for very high temperatures, where the phononic part tends to saturate.\\

\begin{figure}
\includegraphics[scale=1.0]{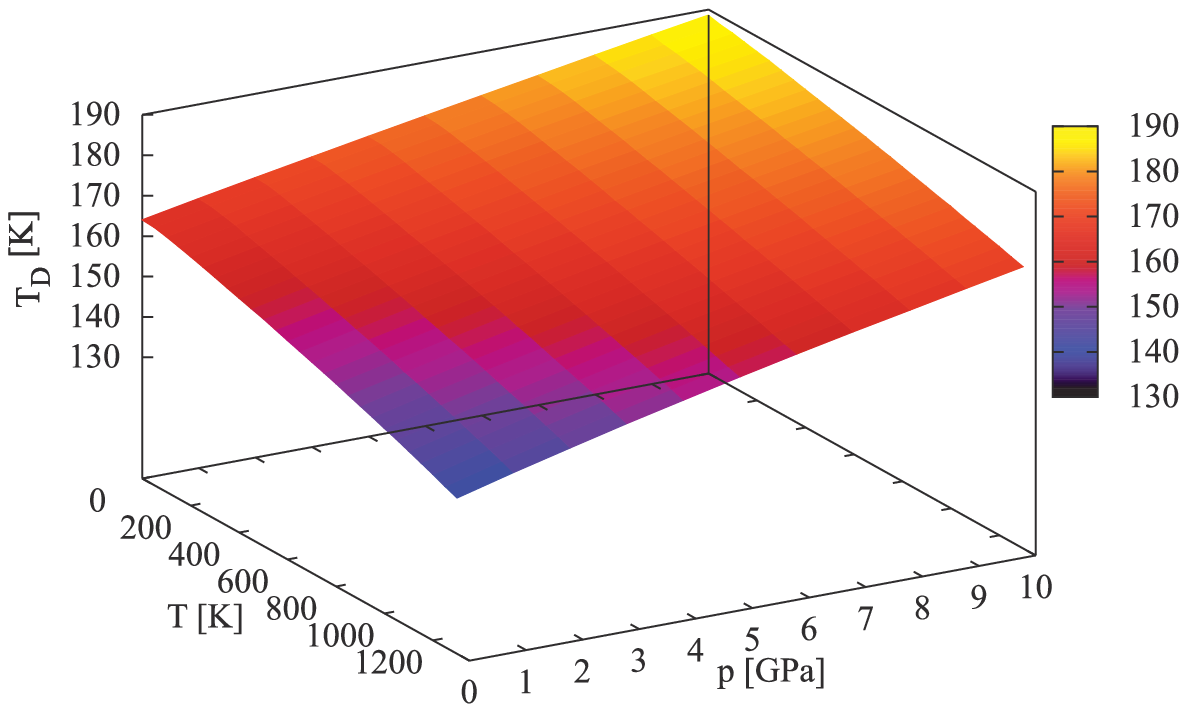}
\caption{(Color online) Debye temperature $T_{\rm D}$ as a function  of pressure $p$ and temperature $T$.}
\label{fig8}
\end{figure}

All calculations in the low-temperature region have been done on the basis of EOS in the form of Eq.~(\ref{eq:eq49}). For higher temperatures, when $T\lesssim 0.16\, T_{\rm D}^0$, the appropriate EOS is given by Eq.~(\ref{eq:eq51}). Fig.~\ref{fig4} presents the thermal expansion coefficient vs. $T$ in the range of high temperatures, limited by the melting temperature of bulk gold ($ T_{\rm M} \approx 1337 $K for $p=0$ \cite{Linde}). A comparison of calculations with the experimental data taken from Ref.~\cite{Anderson} and Ref.~\cite{Yokoo} (after Ref.~\cite{Touloukian}) is made. One can see that present result fits well both sets of experimental data, which have been obtained in different temperature regions. Moreover, for $T \to T_{\rm M}$, the calculated thermal expansion coefficient shows better agreement with the experimental data than, for instance, in Ref.~\cite{Brian}.\\

Regarding compressibility, it should be said that in low-temperature region ($T\lesssim 16$K) it is almost constant ($\kappa_T \approx \kappa_S \approx \kappa _0$) and therefore has not been presented.
Adiabatic compressibility vs. temperature in the high-temperature region is shown in Fig.~\ref{fig5}. The numerical results are compared with the experimental data obtained in Refs.~\cite{Neighbours} and \cite{Yokoo} (after Ref.~\cite{Collard}). A satisfactory agreement between the present theory and experiment can be noted, although near the melting point some differences are more noticeable. The same remark can be made on Fig.~\ref{fig6}, where the specific heat, $C_p$, is presented vs. temperature. In this case the numerical results are compared with the experimental data taken from Refs.~\cite{Anderson} and \cite{Yokoo} (after Ref.~\cite{Barin}). One can see that the agreement between theory and experiment is worse for the highest temperatures, although the difference does not exceed $\approx 3 \%$. In particular, for $T \to T_{\rm M}$, the calculated specific heat is slightly higher than the experimental one. The same tendency has been observed in Ref.~\cite{Yokoo}. The electron contribution to the specific heat is too small to be presented in Fig.~\ref{fig6} as a separate curve. For instance, at temperatures 100 K, 500 K, 1000 K and 1300 K it amounts to $\approx$ 0.3\%, 1.3\%, 2.6\% and 3.5\% of the total specific heat, respectively.\\

The isotherm curve, describing the volume deformation vs. external pressure, is shown in Fig.~\ref{fig7}. The constant temperature amounts to $T=300$ K. The numerical calculations are compared with the experimental data which have been re-calculated from lattice constant measurements \cite {Takemura}. A non-linear decrease of $\varepsilon$ vs. $p$ can be noted, and the negative values of $\varepsilon$ correspond to volume compression. The positive value of $\varepsilon$ for $p=0$, $\varepsilon \approx 0.01$, is connected with thermal expansion in the temperature range from $T=0$ K (where $\varepsilon =0$) up to $T=300$ K. For higher pressures the calculations become less accurate. The similar discrepancy between theory and experiment for high pressures has been observed in Ref.~\cite{Karbasi}.\\

In the last figure (Fig.~\ref{fig8}) the Debye temperature is plotted as a function of two variables: pressure and temperature. The calculations are based on Eq.~\ref{eq:eq8}, whereas $\varepsilon$ is calculated from EOS as a function of $p$ and $T$. For $p=0$ and $T=0$ the Debye temperature starts from the value $T_{\rm D}^0=164$ K and decreases when $T$ increases. The dependence of $T_{\rm D}$ on $p$ is just opposite; the increasing pressure causes increase of the Debye temperature. Such a behaviour allows to deduce that decrease of the Debye temperature can be obtained by increasing the atomic volume. This conclusion is in agreement with the experimental results of Ref.~\cite{Kusaba}, where the Debye temperature was calculated from Debye-Waller factor by an X-ray diffraction method, and a similar dependence has been reported. However, due to approximate expression for the elastic energy, and specific form of Eq.~\ref{eq:eq8}, some of details concerning the Debye temperature behaviour, like an anomaly at 
low temperatures \cite{Gupta,Lynn}, have 
not been reproduced in our calculations.\\

\section{Summary and final conclusions}
 
In the paper  we developed the self-consistent model for thermodynamic description of metallic systems. The idea presented in Ref.~\cite{Balcerzak}, where the Einstein model was combined with the elastic one, has been extended here for the Debye approximation. What is more, the electronic subsystem with its kinetic and exchange energy has been taken into account. The electronic energy has been considered in better approximation than in previous works \cite{Greeff, Souvatzis}. We have shown that the ground state (Fermi) energy, as well as the exchange energy, both of them being volume dependent, can contribute to the electronic pressure.
The regions of low and high temperatures are described by different EOS, however, the whole temperature range $0\le T< T_{\rm M}$ has been covered by these equations. 
Contrary to Birch-Murnaghan equation of state, which presents only an isothermal description \cite{Wallace}, in our EOS all of the variables ($p,V,$ and $T$) are treated equivalently.
The numerical results have been obtained for gold crystal showing satisfactory agreement with the experimental data, as well as with some DFT calculations. The difference between our results and the experimental data does not exceed $\approx 3 \%$ for the best fit of theoretical parameters.\\

The greatest difference between the numerical results and experiment turns out to be near the melting point. A possible source of such inaccuracy is the Debye approximation. This approximation has been used for the sake of simplicity, however, it is rather coarse for the real systems at high temperatures. Another reason is that our starting point for the series expansion of elastic energy with respect to $\varepsilon$ is ($T=0$, $p=0$). 
Thus, our theory is most applicable around the equilibrium point ($p=0,\, T=0,$ and $\varepsilon =0$).
For the case of gold it is quite far from the melting point. 
For instance, in high-pressure equation of state (Birch-Murnaghan) $T_0 \approx 300$ K and $p_0=1$ bar has been assumed as a reference temperature and pressure point.
In our case, expanding the range of pressures up to some extreme values would require higher order terms vs. $\varepsilon$ and new fitting parameters in the elastic potential to be taken into account, as well as other necessary improvements on the presented approach. For instance,
the assumption that the elastic coefficients $B$, $C$, $D$, etc., are constants in the whole temperature and pressure region is only an approximation. It has been shown that the elastic coefficients of gold are, to some extent, pressure \cite{Daniels} and temperature \cite{Neighbours,Garai} dependent. For the above reasons, the application of the method in its present form for the range of extreme pressures would be, in our opinion, rather limited in practice.\\

Similarly to other models leading to EOS, in our calculations we have used only a single variable $\varepsilon$ for description of the volume elastic deformation. However, the approach can be generalized for anisotropic deformations (also including anisotropic external pressures). It should also be mentioned that our considerations are limited to the quasistatic processes and the shock-wave experiments cannot be described within this model.\\

In this paper the numerical calculations have been performed for the case of gold only. The analysis for another metal can be done analogously, whereas the numerical calculations of all thermodynamic properties should be performed simultaneously from one set of fitting parameters. Among these parameters there are characteristics of elastic potential ($C$, $D$, $E$, etc.), as well as other parameters ($q$, $r$) which form an unique set for a given metal. The best fitting of all curves to the experimental data means in practice multiple and time-consuming calculations. However, the numerical calculations and analysis of the properties for other metals is beyond the scope of present paper, which is mainly devoted to the detailed presentation of  theoretical model.\\ 

In spite of the features mentioned above, in our opinion, the presented model can be useful for thermodynamic description of metallic systems. Its advantage follows from the fact that all thermodynamic properties can be found on the basis of one single expression: the Gibbs energy. Hence, the self-consistency of the theory is preserved and all thermodynamic relationships (like Gr\"uneisen equation) are exactly fulfilled. As mentioned above, the model can be further improved when the volume deformation $\varepsilon$ is treated as anisotropic quantity. On the other hand, the vibrational energy can be taken more accurately than in the Debye approximation, for instance, by better modelling of the phononic dispersion relations and density of states.  Also, the electronic energy calculations would benefit from a more realistic model of band structure. For further improvement of the model, electron-phonon interaction might also be taken into account. Of course, 
such improvements will make the model more accurate, but, at the same time, more complicated for the practical use.\\
 

\end{document}